\documentclass[rnote,traditabstract]{aa}

\usepackage{graphicx}
\usepackage{txfonts}
\usepackage{natbib}
\bibpunct{(}{)}{;}{a}{}{,}

\newcommand{\Msun}{\ensuremath{\,{\rm M}_\odot}}                  
\newcommand{\Rsun}{\ensuremath{\,{\rm R}_\odot}}                  
\newcommand{\Teff}{\ensuremath{T_{\rm eff}}}                      
\newcommand{\logg}{\ensuremath{\log g}}                           
\newcommand{\Mjup}{\ensuremath{\,{\rm M}_{\rm Jup}}}              
\newcommand{\Rjup}{\ensuremath{\,{\rm R}_{\rm Jup}}}              
\newcommand{\Teq}{\ensuremath{T_{\rm eq}^{\,\prime}}}             
\newcommand{\safronov}{\ensuremath{\Theta}}                       
\newcommand{\kms}{\,km\,s$^{-1}$}                                 
\newcommand{\ms}{\,m\,s$^{-1}$}                                   
\newcommand{\mss}{\,m\,s$^{-2}$}                                  
\newcommand{\as}{\ensuremath{^{\prime\prime}}}                    
\newcommand{\FeH}{\ensuremath{\left[\frac{\rm Fe}{\rm H}\right]}} 
\newcommand{\pjup}{\ensuremath{\,\rho_{\rm Jup}}}                 
\newcommand{\psun}{\ensuremath{\,\rho_\odot}}                     
\newcommand{\mc}[1]{\multicolumn{2}{c}{#1}}
\newcommand{\mcc}[1]{\multicolumn{3}{c}{#1}}
\newcommand{\er}[3]{\ensuremath{#1^{+#2}_{-#3}}}
\newcommand{\erc}[3]{\mc{\ensuremath{#1^{+#2}_{-#3}}}}

\newcommand{\ermcc}[5]{\mcc{\ensuremath{{#1\,^{+#2}_{-#3}}\,^{+#4}_{-#5}}}}
\newcommand{\reff}[1]{{#1}}                                   

\begin{document} 

\title{A much lower density for the transiting extrasolar planet WASP-7\thanks{Based on data collected by MiNDSTEp with the Danish 1.54m telescope at the ESO La Silla Observatory.}}

\titlerunning{A much lower density for the transiting exoplanet WASP-7}

\author{John Southworth \inst{\ref{inst1}} \and                                               
        M.\ Dominik\thanks{Royal Society University Research Fellow} \inst{\ref{inst2}} \and  
        U.\ G.\ J{\o}rgensen \inst{\ref{inst4},\ref{inst21}} \and                             
        S.\ Rahvar \inst{\ref{inst5}} \and                                                    
        C.\ Snodgrass \inst{\ref{inst6},\ref{inst22}} \and                                    
        K.\ Alsubai \inst{\ref{inst19}} \and                                                  
        V.\ Bozza \inst{\ref{inst8},\ref{inst9},\ref{inst10}} \and                            
        P.\ Browne \inst{\ref{inst2}} \and                                                    
        M.\ Burgdorf \inst{\ref{inst7},\ref{inst20}} \and                                     
        S.\ Calchi Novati \inst{\ref{inst8},\ref{inst10}} \and                                
        P.\ Dodds \inst{\ref{inst2}} \and                                                     
        S.\ Dreizler \inst{\ref{inst13}} \and                                                 
        F.\ Finet \inst{\ref{inst11}} \and                                                    
        T.\ Gerner \inst{\ref{inst16}} \and                                                   
        S.\ Hardis \inst{\ref{inst4},\ref{inst21}} \and                                       
        K.\ Harps{\o}e \inst{\ref{inst4},\ref{inst21}} \and                                   
        C.\ Hellier \inst{\ref{inst1}} \and                                                   
        T.\ C.\ Hinse \inst{\ref{inst4},\ref{inst3}} \and                                     
        M.\ Hundertmark \inst{\ref{inst13}} \and                                              
        N.\ Kains \inst{\ref{inst2},\ref{inst14}} \and                                        
        E.\ Kerins \inst{\ref{inst15}} \and                                                   
        C.\ Liebig \inst{\ref{inst2}} \and                                                    
        L.\ Mancini \inst{\ref{inst8},\ref{inst10},\ref{inst17}} \and                         
        M.\ Mathiasen \inst{\ref{inst4}} \and                                                 
        M.\ T.\ Penny \inst{\ref{inst15}} \and                                                
        S.\ Proft \inst{\ref{inst16}} \and                                                    
        D.\ Ricci \inst{\ref{inst11}} \and                                                    
        K.\ Sahu \inst{\ref{inst18}} \and                                                     
        G.\ Scarpetta \inst{\ref{inst8},\ref{inst9},\ref{inst10}} \and                        
        S.\ Sch\"afer \inst{\ref{inst13}} \and                                                
        F.\ Sch\"onebeck \inst{\ref{inst16}} \and                                             
        J.\ Surdej \inst{\ref{inst11}}                                                        
        }


\authorrunning{John Southworth et al.}

\institute{Astrophysics Group, Keele University, Staffordshire, ST5 5BG, UK \ \ \ \ \ \email{jkt@astro.keele.ac.uk} \label{inst1} \and
      SUPA, University of St Andrews, School of Physics \& Astronomy, North Haugh, St Andrews, KY16 9SS, UK \label{inst2} \and
      Niels Bohr Institute, University of Copenhagen, Juliane Maries vej 30, 2100 Copenhagen \O, Denmark \label{inst4} \and
      Centre for Star and Planet Formation, Geological Museum, {\O}ster Voldgade 5, 1350 Copenhagen, Denmark \label{inst21} \and
      Department of Physics, Sharif University of Technology, P.\,O.\,Box 11155-9161 Tehran, Iran \label{inst5} \and
      Max-Planck-Institute for Solar System Research, Max-Planck Str.\ 2, 37191 Katlenburg-Lindau, Germany \label{inst6} \and
      European Southern Observatory, Casilla 19001, Santiago 19, Chile \label{inst22} \and
      Qatar Foundation, Doha, Qatar \label{inst19} \and
      Dipartimento di Fisica ``E. R. Caianiello'', Universit\`a di Salerno, Via Ponte Don Melillo, 84084-Fisciano (SA), Italy \label{inst8} \and
      Deutsches SOFIA Institut, Universitaet Stuttgart, Pfaffenwaldring 31, 70569 Stuttgart, Germany \label{inst7} \and
      SOFIA Science Center, NASA Ames Research Center, Mail Stop N211-3, Moffett Field CA 94035, USA \label{inst20} \and
      Istituto Nazionale di Fisica Nucleare, Sezione di Napoli, Napoli, Italy \label{inst9} \and
      Istituto Internazionale per gli Alti Studi Scientifici (IIASS), 84019 Vietri Sul Mare (SA), Italy \label{inst10} \and
      Institut f\"ur Astrophysik, Georg-August-Universit\"at G\"ottingen, Friedrich-Hund-Platz 1, 37077 G\"ottingen, Germany \label{inst13} \and
      European Southern Observatory, Karl-Schwarzschild-Stra{\ss}e 2, 85748 Garching bei M\"unchen, Germany \label{inst14} \and
      Institut d'Astrophysique et de G\'eophysique, Universit\'e de Li\`ege, 4000 Li\`ege, Belgium \label{inst11} \and
      Astronomisches Rechen-Institut, Zentrum f\"ur Astronomie, Universit\"at Heidelberg, M\"onchhofstra{\ss}e 12-14, 69120 Heidelberg \label{inst16} \and 
      Armagh Observatory, College Hill, Armagh, BT61 9DG, Northern Ireland, UK \label{inst3} \and
      Jodrell Bank Centre for Astrophysics, University of Manchester, Oxford Road, Manchester, M13 9PL, UK \label{inst15} \and   
      Dipartimento di Ingegneria, Universit\`a del Sannio, Corso Garibaldi 107, 82100-Benevento, Italy \label{inst17} \and
      Space Telescope Science Institute, 3700 San Martin Drive, Baltimore, MD. 21218, USA \label{inst18}
      }


\abstract{We present the first high-precision photometry of the transiting extrasolar planetary system WASP-7, obtained using telescope defocussing techniques and reaching a scatter of 0.68\,mmag per point. We find that the transit depth is greater and that the host star is more evolved than previously thought. The planet has a significantly larger radius ($1.330 \pm 0.093$\Rjup\ versus \er{0.915}{0.046}{0.040}\Rjup) and much lower density ($0.41 \pm 0.10$\pjup\ versus \er{1.26}{0.25}{0.21}\pjup) and surface gravity ($13.4 \pm 2.6$\mss\ versus \er{26.4}{4.4}{4.0}\mss) than previous measurements showed. Based on the revised properties it is no longer an outlier in planetary mass--radius and period--gravity diagrams. \reff{We also obtain a more precise transit ephemeris for the WASP-7 system.}
}


\maketitle 

\section{Introduction}

The transiting extrasolar planet (TEP) \object{WASP-7}\,b was discovered by the WASP consortium \citep[][hereafter H09]{Hellier+09apj} through the detection of transits in front of its F5\,V parent star. It is a challenging target for acquiring high-precision transit photometry, due to the brightness of the parent star ($V = 9.5$), the paucity of good nearby comparison stars, the transit duration (3.8\,hr), and the orbital period ($4.95$\,d) which is both comparatively long and close to an integer number of days. The characterisation of WASP-7 therefore relied upon the photometry obtained by the WASP-South telescope \citep{Pollacco+06pasp}.

The relatively large scatter in the discovery data meant that the transit shape was poorly delineated. Because of this, the analysis by H09 included an additional constraint in the form of a main sequence mass--radius relation for the host star \citep[e.g.][]{Anderson+10apj}. \reff{The radius, surface gravity and density of the planet resulting from their analysis are $R_{\rm b} = \er{0.91}{0.046}{0.040}$\Rjup, $g_{\rm b} = \er{26.4}{4.4}{4.0}$\mss\ and $\rho_{\rm b} = \er{1.26}{0.25}{0.21}$\pjup, respectively. These values placed WASP-7\,b in an outlier position in the mass--radius diagram of TEPs, having one of the largest densities within the main planet population (masses $\la$2\Mjup). This was interpreted by H09 as evidence that WASP-7\,b has a massive heavy-element core.}

In this work we present the first follow-up photometric observations obtained for WASP-7. The high precision of our observations (0.68\,mmag scatter) allows us to obtain the physical properties of the transiting system without needing to impose any constraints on the parameters of the parent star. We find a substantially larger radius, \reff{and therefore a lower density and surface gravity}. We also greatly improve the orbital ephemeris for the system, so transit midpoints in the 2011 observing season can be predicted to within 45\,s instead of 27\,min.


\section{Observations and data reduction}

\begin{table*} \centering \caption{\label{tab:lcfits} Parameters of the {\sc jktebop}
best fits of the light curve of WASP-7, using different approaches to limb darkening (LD).}
\begin{tabular}{l r@{\,$\pm$\,}l r@{\,$\pm$\,}l r@{\,$\pm$\,}l r@{\,$\pm$\,}l r@{\,$\pm$\,}l}
\hline \hline
\                     &      \mc{Linear LD law}   &    \mc{Quadratic LD law}  &   \mc{Square-root LD law} &   \mc{Logarithmic LD law} &     \mc{Cubic LD law}     \\
\hline
\multicolumn{11}{l}{All LD coefficients fixed:} \\[1pt]
$r_{\rm A}+r_{\rm b}$ & 0.1255       & 0.0052     & 0.1180       & 0.0049     & 0.1194       & 0.0048     & 0.1184       & 0.0051     & 0.1223       & 0.0046     \\
$k$                   & 0.09598      & 0.00089    & 0.09480      & 0.00075    & 0.09524      & 0.00075    & 0.09475      & 0.00082    & 0.09633      & 0.00060    \\
$i$ ($^\circ$)        & 86.61        &  0.52      & 87.43        &  0.59      & 87.23        &  0.56      & 87.39        &  0.63      & 86.80        &  0.47      \\
$u_{\rm A}$           &      \mc{ 0.43 fixed}     &      \mc{ 0.20 fixed}     &      \mc{ 0.00 fixed}     &      \mc{ 0.55 fixed}     &      \mc{ 0.20 fixed}     \\
$v_{\rm A}$           &            \mc{ }         &      \mc{ 0.30 fixed}     &      \mc{ 0.60 fixed}     &      \mc{ 0.25 fixed}     &      \mc{ 0.15 fixed}     \\[2pt]
$r_{\rm A}$           & 0.1145       & 0.0047     & 0.1078       & 0.0044     & 0.1090       & 0.0044     & 0.1081       & 0.0046     & 0.1116       & 0.0041     \\
$r_{\rm b}$           & 0.01099      & 0.00053    & 0.01022      & 0.00048    & 0.01038      & 0.00048    & 0.01025      & 0.00051    & 0.01075      & 0.00045    \\
$\sigma$ (mmag)       &        \mc{ 0.7136}       &        \mc{ 0.6977}       &        \mc{ 0.6946}       &        \mc{ 0.7015}       &        \mc{ 0.6828}       \\
\hline
\multicolumn{11}{l}{Fitting for the linear LD coefficient and perturbing the nonlinear LD coefficient:} \\[1pt]
$r_{\rm A}+r_{\rm b}$ & 0.1224       & 0.0043     & 0.1192       & 0.0047     & 0.1227       & 0.0046     & 0.1221       & 0.0047     & 0.1200       & 0.0044     \\
$k$                   & 0.09675      & 0.00069    & 0.09587      & 0.00072    & 0.09646      & 0.00065    & 0.09628      & 0.00069    & 0.09623      & 0.00071    \\
$i$ ($^\circ$)        & 86.77        &  0.44      & 87.15        &  0.55      & 86.77        &  0.48      & 86.83        &  0.49      & 87.05        &  0.49      \\
$u_{\rm A}$           &  0.245       &  0.047     &  0.059       &  0.064     & -0.134       &  0.059     &  0.387       &  0.064     &  0.190       &  0.051     \\
$v_{\rm A}$           &            \mc{ }         &     \mc{ 0.30 perturbed}  &     \mc{ 0.60 perturbed}  &     \mc{ 0.25 perturbed}  &     \mc{ 0.15 perturbed}  \\[2pt]
$r_{\rm A}$           & 0.1116       & 0.0039     & 0.1088       & 0.0043     & 0.1119       & 0.0042     & 0.1113       & 0.0043     & 0.1095       & 0.0039     \\
$r_{\rm b}$           & 0.01080      & 0.00043    & 0.01043      & 0.00047    & 0.01079      & 0.00045    & 0.01072      & 0.00046    & 0.01054      & 0.00042    \\
$\sigma$ (mmag)       &        \mc{ 0.6841}       &        \mc{ 0.6829}       &        \mc{ 0.6826}       &        \mc{ 0.6825}       &        \mc{ 0.6827}       \\
\hline
\multicolumn{11}{l}{Fitting for both LD coefficients:} \\[1pt]
$r_{\rm A}+r_{\rm b}$ &            \mc{ }         & 0.1216       & 0.0050     & 0.1199       & 0.0049     & 0.1221       & 0.0054     & 0.1196       & 0.0050     \\
$k$                   &            \mc{ }         & 0.09583      & 0.00115    & 0.09537      & 0.00145    & 0.09575      & 0.00148    & 0.09527      & 0.00151    \\
$i$ ($^\circ$)        &            \mc{ }         & 86.91        &  0.56      & 87.15        &  0.60      & 86.89        &  0.63      & 87.18        &  0.63      \\
$u_{\rm A}$           &            \mc{ }         & -0.021       &  0.256     & -0.866       &  0.979     &  0.578       &  0.391     &  0.066       &  0.131     \\
$v_{\rm A}$           &            \mc{ }         & 0.42         & 0.41       & 1.83         & 1.60       & 0.56         & 0.60       & 0.54         & 0.45       \\[2pt]
$r_{\rm A}$           &            \mc{ }         & 0.1110       & 0.0045     & 0.1095       & 0.0045     & 0.1114       & 0.0049     & 0.1092       & 0.0045     \\
$r_{\rm b}$           &            \mc{ }         & 0.01064      & 0.00051    & 0.01044      & 0.00051    & 0.01067      & 0.00055    & 0.01040      & 0.00052    \\
$\sigma$ (mmag)       &            \mc{ }         &        \mc{ 0.6826}       &        \mc{ 0.6818}       &        \mc{ 0.6820}       &        \mc{ 0.6817}       \\
\hline \end{tabular} \end{table*}

WASP-7 was observed from 00:14 to 07:18 UT on the date 2010/09/07. We used the 1.54\,m Danish telescope at ESO La Silla, equipped with the DFOSC focal-reducing CCD imager. A total of 217 integrations were obtained through a Gunn $I$ filter (ESO filter \#425) and with an exposure time of 60\,s. The CCD was windowed down in order to reduce readout time, resulting in a dead time of 50\,s between individual observations. The telescope was defocussed to a point spread function diameter of 38\,pixels (15\as) in order to average out flat-fielding noise and avoid saturating the CCD.

The data were reduced using a pipeline which implements the {\sc daophot} aperture photometry routine \citep{Stetson87pasp}. The images were debiassed and flat-fielded, and slight pointing variations were detected and accounted for by cross-correlating against a reference image. A differential-magnitude light curve was constructed by simultaneously fitting a straight line to the out-of-transit observations plus weights to the reference stars used to create the ensemble comparison star. We find that the sizes of the software apertures and the choice of comparison stars has very little effect on the shape of the observed transit. For our final photometry we adopt the light curve which displays the lowest scatter with respect to a fitted transit model (see below). The root-mean-square (rms) of the scatter in this light curve is 0.68\,mmag per point. For further details on the approach used to obtain and reduce the data see \citet{Me+09mn,Me+10mn}.


\section{Light curve analysis}

The light curve of WASP-7 was modelled using the {\sc jktebop} code\footnote{{\sc jktebop} is written in {\sc fortran77} and the source code is available at {\tt http://www.astro.keele.ac.uk/$\sim$jkt/codes/jktebop.html}}, which in turn is based on the {\sc ebop} program \citep{PopperEtzel81aj,NelsonDavis72apj}. The main parameters of the fit were the sum of the fractional radii of the star and planet, $r_{\rm A} + r_{\rm b} = \frac{R_{\rm A}+R_{\rm b}}{a}$ where $a$ is the semimajor axis, their ratio, $k = \frac{r_{\rm b}}{r_{\rm A}}$ and the orbital inclination, $i$. After a preliminary fit was obtained, the measurement errors were rescaled to give a reduced $\chi^2$ of $1.0$. In order to provide an accurate measure of the orbital period we included the time of mid-transit quoted by H09, \reff{using the approach given by \citet{Me++07aa}.} This results in an orbital ephemeris of:
$$ T_0 = {\rm BJD(TDB)} \,\, 2\,455\,446.63493 (30) \, + \, 4.9546416 (35) \times E $$
where $E$ is the number of orbital cycles after the reference epoch and the bracketed quantities denote the uncertainty in the final digit of the preceding number. This ephemeris is on the \reff{BJD(TDB)} timescale \citep{Eastman++10pasp}. The timestamps in our data were manually verified to be correct to within $\pm$2\,s.

\begin{table} \centering \caption{\label{tab:lcfinal} Final parameters of
the fits to the light curve of WASP-7, compared to the results from H09.}
\begin{tabular}{l r@{\,$\pm$\,}l c} \hline \hline
\                     &  \mc{This work}   & H09                 \\
\hline
$r_{\rm A}+r_{\rm b}$ & 0.1207  & 0.0068  & 0.1001              \\
$k$                   & 0.0956  & 0.0016  & 0.0761 $\pm$ 0.0008 \\
$i$ ($^\circ$)        & 87.03   & 0.93    & \er{89.6}{0.4}{0.9} \\
$r_{\rm A}$           & 0.1102  & 0.0061  & 0.0930              \\
$r_{\rm b}$           & 0.01053 & 0.00070 & 0.00708             \\
\hline \end{tabular} \end{table}

\begin{table*} \begin{center}
\caption{\label{tab:models} Derived physical properties of
WASP-7 using five different sets of theoretical stellar models.}
\begin{tabular}{l r@{\,$\pm$\,}l r@{\,$\pm$\,}l r@{\,$\pm$\,}l r@{\,$\pm$\,}l r@{\,$\pm$\,}l r@{\,$\pm$\,}l}
\hline \hline
\ & \mc{({\it Claret} models)} & \mc{({\it Y$^2$} models)} & \mc{({\it Teramo} models)} & \mc{({\it VRSS} models)} & \mc{({\it DSEP} models)} \\
\hline
$K_{\rm b}$     (\kms) & 135.6   &   2.1     & 135.6   &   1.9     & 135.1   &   2.2     & 135.4   &   1.9     & 134.4   &   2.2     \\
$M_{\rm A}$    (\Msun) & 1.285   & 0.059     & 1.287   & 0.054     & 1.273   & 0.061     & 1.280   & 0.052     & 1.254   & 0.061     \\
$R_{\rm A}$    (\Rsun) & 1.436   & 0.092     & 1.436   & 0.092     & 1.431   & 0.092     & 1.434   & 0.090     & 1.424   & 0.092     \\
$\log g_{\rm A}$ (cgs) & 4.233   & 0.047     & 4.233   & 0.047     & 4.231   & 0.047     & 4.232   & 0.047     & 4.229   & 0.047     \\[2pt]
$M_{\rm b}$    (\Mjup) & 0.96    & 0.13      & 0.96    & 0.13      & 0.96    & 0.13      & 0.96    & 0.13      & 0.95    & 0.13      \\
$R_{\rm b}$    (\Rjup) & 1.333   & 0.093     & 1.334   & 0.093     & 1.329   & 0.093     & 1.331   & 0.092     & 1.322   & 0.092     \\
$\rho_{\rm b}$ (\pjup) & 0.407   & 0.101     & 0.406   & 0.101     & 0.408   & 0.101     & 0.407   & 0.101     & 0.410   & 0.102     \\
\safronov\             & 0.069   & 0.011     & 0.069   & 0.010     & 0.070   & 0.011     & 0.070   & 0.011     & 0.070   & 0.011     \\[2pt]
$a$               (AU) & 0.06185 & 0.00095   & 0.06188 & 0.00086   & 0.06166 & 0.00098   & 0.06177 & 0.00084   & 0.06134 & 0.00100   \\
Age              (Gyr) & \erc{2.5}{0.5}{1.0} & \erc{2.5}{0.2}{1.0} & \erc{2.3}{0.6}{0.8} & \erc{2.0}{0.7}{0.6} & \erc{2.7}{0.8}{0.7} \\
\hline \end{tabular} \end{center}
\end{table*}

Limb darkening (LD) was included and its uncertainty accounted for using five different functional laws, and solutions were obtained for the cases when both coefficients were fixed to theoretical values, the linear coefficient was fitted for, and both coefficients were fitted for. \reff{Theoretical values were calculated for the measured effective temperature and surface gravity of the star by linearly interpolating within the tables of \citet{Vanhamme93aj} and \citet{Claret00aa}.} The results for each of these modelling runs are given in Table\,\ref{tab:lcfits}. \reff{We find that the results for different LD laws are generally in good agreement, although the amount of LD is lower than theoretically predicted.}

\begin{figure} \includegraphics[width=\columnwidth,angle=0]{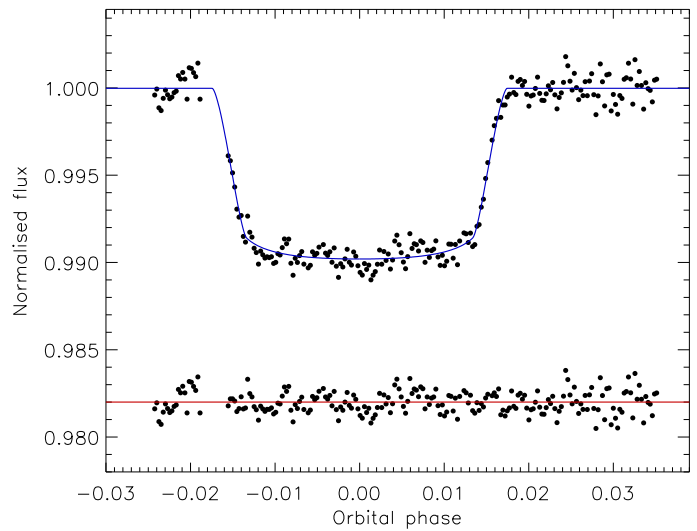}
\caption{\label{fig:lcfit} Phased light curve of WASP-7 compared to the best
fit found using {\sc jktebop}. The residuals of the fits are plotted at the
base of the figure, offset from zero.} \end{figure}

Uncertainties in the fitted parameters were obtained using Monte Carlo and residual-permutation simulations \citep{Me08mn,Jenkins++02apj}. We are able to reject the solutions with all LD coefficients fixed to theoretical values, as they provide a clearly worse fit to the data. The solutions where two LD coefficients are fitted are marginally the best ones, so for the final parameters we adopt their mean values from these four solutions. We adopt the parameter uncertainties from the residual permutation algorithm, which are roughly 1.5 times larger than the Monte Carlo alternatives. This implies that correlated noise is significant in our data, as expected from visual inspection of the plot of the best fit (Fig.\,\ref{fig:lcfit}). The final parameter values and uncertainties are given in Table\,\ref{tab:lcfinal}. Compared to those from H09, we find a solution with a lower orbital inclination and larger fractional radius for both the star and the planet. Our results were obtained using the methods of \citet{Me08mn} and are homogeneous with the results in that work.

\section{The physical properties of WASP-7}                                                                                         \label{sec:absdim}

The light curve parameters alone do not allow the physical properties of the WASP-7 system to be calculated. Additional observed properties are available from H09: the star's orbital velocity amplitude ($K_{\rm A} = 97 \pm 13$\ms), effective temperature ($\Teff = 6400 \pm 100$\,K), surface gravity ($\logg = 4.3 \pm 0.2$) and metallicity ($\FeH = 0.0 \pm 0.1$). Of these, the surface gravity is quite imprecise so is only useful as a consistency check.

In order to calculate the physical properties of WASP-7 we use the same approach as in \citet{Me09mn}. Firstly, we adopt tabulated predictions from theoretical models of low-mass stars. Secondly, we guess an initial value of the velocity amplitude of the planet, $K_{\rm b}$. We then use the known $r_{\rm A}$, $r_{\rm b}$, $i$, $K_{\rm A}$ and orbital period to calculate the physical properties of the system. The observed \Teff\ and calculated $R_{\rm A}$ are then compared to the model-predicted values for a star of the calculated mass ($M_{\rm A}$), and $K_{\rm b}$ is adjusted until the best match is found. This is performed for a grid of ages covering 0.1 to 20 Gyr in 0.1\,Gyr steps, by which the overall best physical properties and age are found. The uncertainty in each of the input parameters is propagated by repeating this procedure with the parameter perturbed, resulting in a detailed error budget for every output parameter.

The above procedure is performed five times, using independent sets of theoretical model predictions. These are: {\it Claret} \citep{Claret04aa}, {\it Y$^2$} \citep{Demarque+04apjs}, {\it Teramo} \citep{Pietrinferni+04apj}, {\it VRSS} \citep{Vandenberg++06apjs} and {\it DSEP} \citep{Dotter+08apjs}. The final value for each of the physical properties is taken to be the unweighted mean of the values found using the five different model sets. The random error is taken to be the largest of the individal errors found by the perturbation analysis. The systematic error is the standard deviation of the values from the five different model sets. The results for each model set are given in Table\,\ref{tab:models}. The final physical properties of WASP-7 are collected in Table\,\ref{tab:final} and compared to those found by H09. \reff{In this table the quantity $\Teq$ is the equilibrium temperature of the planet excluding the energy redistribution factor \citep{Me09mn} and \safronov\ is the \citet{Safronov72} number.} We find that the star is slightly evolved, allowing a reasonable estimate of its age ($2.4 \pm 1.0$\,Gyr).


\section{Summary and conclusions}

\begin{table} \begin{center}
\caption{\label{tab:final} Final physical properties of WASP-7 compared to
those found by H09. The second errorbars represent systematic errors.}
\begin{tabular}{l r@{\,$\pm$\,}c@{\,$\pm$\,}l r@{\,$\pm$\,}l} \hline \hline
\                      & \mcc{This work (final)}       & \mc{H09}                   \\
\hline
$M_{\rm A}$    (\Msun) & 1.276  & 0.061  & 0.022       & \erc{1.28}{0.09}{0.19}     \\
$R_{\rm A}$    (\Rsun) & 1.432  & 0.092  & 0.008       & \erc{1.236}{0.059}{0.046}  \\
$\log g_{\rm A}$ (cgs) & 4.232  & 0.047  & 0.003       & \erc{4.363}{0.010}{0.047}  \\
$\rho_{\rm A}$ (\psun) & \mcc{$0.434 \pm 0.074$}       & \mc{ }                     \\[2pt]
$M_{\rm b}$    (\Mjup) & 0.96   & 0.13   & 0.01        & \erc{0.96}{0.12}{0.18}     \\
$R_{\rm b}$    (\Rjup) & 1.330  & 0.093  & 0.008       & \erc{0.915}{0.046}{0.040}  \\
$g_{\rm b}$      (\ms) & \mcc{$13.4 \pm 2.6$}          & \erc{26.4}{4.4}{4.0}       \\
$\rho_{\rm b}$ (\pjup) & \mcc{$0.41 \pm 0.10$}         & \erc{1.26}{0.25}{0.21}     \\[2pt]
\Teq\              (K) &   \mcc{$1487 \pm 48$}         & \erc{1379}{35}{23}         \\
\safronov\             & 0.070  & 0.011  & 0.000       & \mc{ }                     \\
$a$               (AU) & 0.0617 & 0.0010 & 0.0004      &\erc{0.0618}{0.0014}{0.0033}\\
Age              (Gyr) &\ermcc{2.4}{0.8}{1.0}{0.3}{0.4}& \mc{ }                     \\
\hline \end{tabular} \end{center} \end{table}

\begin{figure} \includegraphics[width=\columnwidth,angle=0]{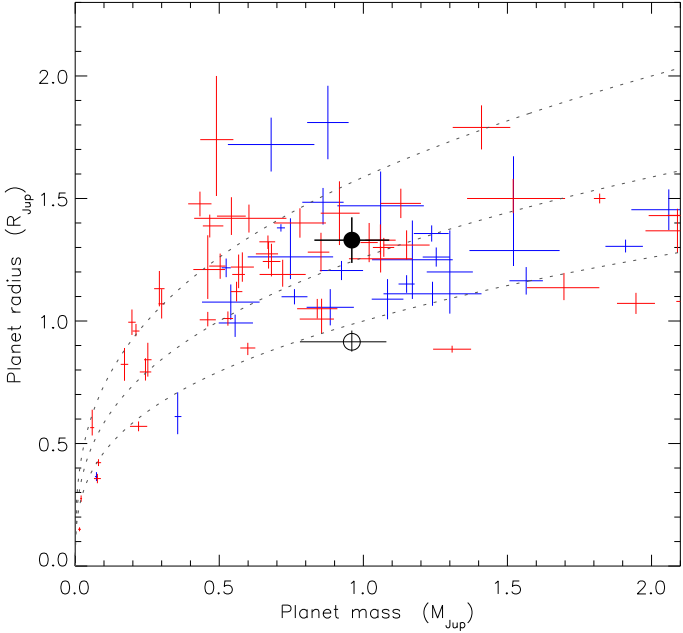}
\caption{\label{fig:m2m3} Plot of the masses and radii of the known
TEPs. The blue symbols denote values from the homogeneous analysis of
\citet{Me10mn} and the red symbols results for the other known TEPs.
WASP-7 is shown in black with an open circle (H09) and a filled circle
(this work). Grey dotted lines show where density is 1.0, 0.5 and 0.25
\pjup.} \end{figure}

\begin{figure} \includegraphics[width=\columnwidth,angle=0]{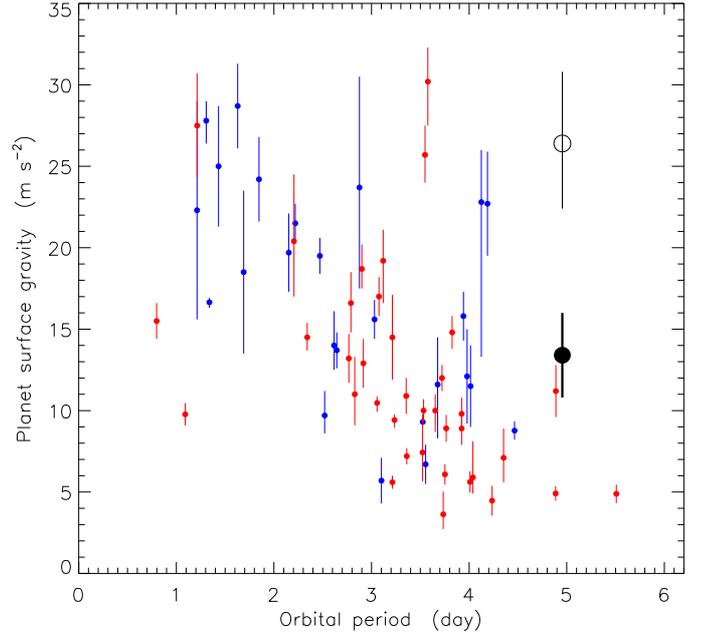}
\caption{\label{fig:pg2} Plot of the orbital periods and surface gravities
of the known TEPs. The symbols are the same as for Fig.\,\ref{fig:m2m3}.}
\end{figure}

We have observed a transit of the WASP-7 system using defocussed photometry techniques, resulting in a light curve with an rms scatter of only 0.68\,mmag. These data were modelled with the {\sc jktebop} code, and the results used to determine the physical properties of the planet and its host star. The higher quality of our photometry has allowed us to measure the system properties without constraining the host star to follow a main-sequence mass-radius relation. Compared to H09, we find both a larger radius for the host star and a larger ratio of the radii of the planet and star. This results in a sizeable increase in the planetary radius, from $\er{0.915}{0.046}{0.040}$\Rjup\ to $1.330 \pm 0.093$\Rjup, which in turn means a lower surface gravity and mean density.

We have collected literature measurements of the physical properties of the 96 known TEPs as of 2010/11/18. For 30 of these objects we used the results of the homogeneous analysis performed by \citet{Me10mn}. In Fig.\,\ref{fig:m2m3} we plot their masses and radii, plus the values from WASP-7 found in this work (filled circle) and from H09 (open circle). The H09 results for WASP-7\,b indicated that it was one of the densest known TEPs below $\sim$2\Mjup, along with the more recently-discovered planets CoRoT-13\,b \citep{Cabrera+10aa} and HAT-P-15\,b \citep{Kovacs+10apj}. The outlier status has now been lifted: our results place WASP-7\,b in a well-populated part of the mass--radius diagram and demonstrate that high-quality data is required to obtain reliable measurements of the properties of TEPs.

Fig.\,\ref{fig:pg2} shows the surface gravities of the known TEPs as a function of their orbital periods. The existing planet population shows an inverse correlation between period and surface gravity \citep{Me++07mn}, at least for the dominant population with periods $\la$10\,d and masses $\la$3\Mjup\ \reff{(see also \citealt{Fressin++09aa})}. The revised properties of WASP-7\,b move it from outlier status to within the sprawl of parameter space occupied by the general planet population.

The theoretical models of irradiated gas giant planets \citet{Bodenheimer++03apj}, \citet{Fortney++07apj} and \citet{Baraffe++08aa} predict radii of no more than 1.16\Rjup\ for a 1.0\Mjup\ planet with a range of chemical compositions and core masses, and without an arbitrary additional heating source. Our upward revision of the radius of WASP-7\,b means that it no longer matches the predictions of these models at the 2$\sigma$ level. \reff{These conclusions could be strengthened by the provision of more radial velocity measurements as well as the acquisition of further photometric observations.}




\begin{acknowledgements}
JS acknowledges funding from STFC in the form of an Advanced Fellowship. We thank the referee for comments which helped to improve the paper. Astronomical research at Armagh Observatory is funded by the Department of Culture, Arts \& Leisure (DCAL). J\,Surdej, DR (boursier FRIA) and FF acknowledge support from the Communaut\'e fran\c{c}aise de Belgique - Actions de recherche concert\'ees - Acad\'emie Wallonie-Europe.
\end{acknowledgements}


\bibliographystyle{aa}

\end{document}